\documentclass[pdflatex,sn-basic]{sn-jnl}

\usepackage{graphicx}%
\usepackage{multirow}%
\usepackage{amsmath,amssymb,amsfonts}%
\usepackage{amsthm}%
\usepackage{mathrsfs}%
\usepackage[title]{appendix}%
\usepackage{xcolor}%
\usepackage{textcomp}%
\usepackage{manyfoot}%
\usepackage{booktabs}%
\usepackage{algorithm}%
\usepackage{algorithmicx}%
\usepackage{algpseudocode}%
\usepackage{listings}%

\theoremstyle{thmstyleone}%
\theoremstyle{thmstyletwo}%

\theoremstyle{thmstylethree}%

\begin{document}

\title[Article Title]{A nomenclature for individual exocomets}

\author[1]{\fnm{Alain} \sur{Lecavelier des Etangs}}\email{lecaveli@iap.fr}

\author*[2]{\fnm{Paul} \sur{Str{\o}m}}\email{paul@strom.ac}

\author[3,4]{\fnm{Darryl Z.}\sur{Seligman}}\email{dzs@msu.edu}

\affil[1]{\orgdiv{Institut d’astrophysique de Paris, CNRS}, \orgname{Sorbonne Université}, \orgaddress{\street{98$^{\rm bis}$~boulevard Arago}, \city{Paris}, \postcode{75014}, \country{France}}}

\affil[2]{\orgdiv{Department of Physics}, \orgname{University of Warwick}, \orgaddress{\street{Gibbet Hill Road}, \city{Coventry}, \postcode{CV4 7AL}, \country{U.K}}}

\affil[3]{\orgdiv{Department of Physics and Astronomy}, \orgname{Michigan State University}, \city{East Lansing}, \postcode{48824}, \state{MI}, \country{USA}}
\affil[4]{\orgdiv{NSF Astronomy and Astrophysics Postdoctoral Fellow}}

\abstract{With recent observational advancements, exocomet studies have entered a new era: we have moved from an epoch where exocomets' signatures were analysed to identify their true nature to a new epoch where individual exocomets' detections are studied in detail to characterise the observed bodies. In this context, as with other astronomical objects such as exoplanets and minor bodies in the solar system, a nomenclature system is needed to uniquely identify any observed exocometary body.
Here, after outlining the purpose of a nomenclature and its required characteristics, we review the information on exocomets that can be included in the nomenclature. Finally, we propose an exocomet nomenclature scheme to identify the individual exocomets that have been discovered or yet to be discovered. Examples of nomenclature names of a few archetypal exocomets are provided.  
}


\maketitle

\section{Introduction}\label{sec1}

The study of exocomets has progressed significantly with advancements in observational capabilities. Initially, research focused on detecting and analysing the signatures of exocomets to understand their general nature. However, we have now transitioned to an era where individual exocomet detections are examined in detail, allowing for the characterisation of these bodies. This shift necessitates a systematic approach to designating the studied exocomets, similar to the established nomenclature for exoplanets and minor bodies within our Solar System. In this paper, we explore the purpose and essential features of an exocomet nomenclature system and propose a comprehensive framework for uniquely identifying these celestial bodies.

\section{The nomenclature principles}

\subsection{The purpose of the nomenclature}


By principle, the fundamental purpose of a nomenclature system is to uniquely designate individual objects within a specific set of objects. Within a given nomenclature, each object is expected to have a unique designation.

However, for a given class of astrophysical objects --- stars, planets, asteroids, comets to name a few --- the multiplicity of adopted nomenclatures often leads to several and in some cases tens of possible, different designations for the same object. On the other hand, a practically useful nomenclature must be sufficiently specific such that a single designation cannot describe different objects.

Finally, we note the distinction between a nomenclature designation and a ``name''. For the purposes of this manuscript, we define a ``name'' as a designation built with a single word or a simple composition of a very few words. For instance, the brightest star in the sky has a name ``Sirius", which is different from its numerous nomenclature designations, including Alpha Canis Majoris, $\alpha$~CMa, HR\,2491 or HD\,48915.

\subsection{Information coded in the nomenclature}

To construct a nomenclature scheme, we must address the question of what kind of information should be conveyed by the nomenclature designation of a given object. In this section we enumerate potential information that could be used to construct the nomenclature scheme. The final selection of information included in the proposed nomenclature will be presented later in Sect.~\ref{Proposed naming convention}. 

\subsubsection{Name of the star}
\label{Name of the star}

For objects in extrasolar systems, a key demographic feature to be included in the nomenclature is the designation of the system itself. Incorporating the name of the host star system is already implemented in the nomenclature of exoplanets. Typical exoplanet designations begin with the catalog name of the host star, such as 51~Peg\,b or HD~209458\,b. 

In this paper we propose a nomenclature scheme where the catalog name of the star designates the extrasolar planetary system harboring the considered exocomet. 
A direct and intentional consequence of this choice is that an exocomet is --- by construction --- bound to and orbiting around a central stellar object. Therefore, this scheme excludes free-floating or interstellar comets under the definition of exocomets. 
Moreover, this approach aligns with the definition of an exoplanet, which requires it to orbit a star, a brown dwarf or a stellar remnant \citep{Lecavelier_2022}. This proposed system is not unprecedented; for example, interstellar comets observed traversing the Solar System have already had comet-like nomenclature schemes applied to them. These comets are designated with a number followed by {\tt I} (for interstellar), such as 1I/'Oumuamua \citep{Williams17} and 2I/Borisov.  

 This scheme leads to the necessity of a choice regarding the star's catalogue name to be used in the nomenclature. A similar issue has occurred for exoplanets, because there are no specific rules for exoplanets and in rare circumstances multiple designations have been applied to the same object (for example for WASP-11\,b, which is the same planet as HAT-P-10\,b, or WASP-33\,b which is also designated by HD15082\,b). 
 To address that issue, we advise to restrict the number of catalogs to be used and to prioritise the use of the most straightforward and shortest designation. 
 We propose to use the following designations in the strict following order of priority:
 
 \begin{enumerate}
     \item The Bayer's designation (like $\beta$\,Pic),
     \item The Flamsteed Catalog (like 51~Peg)
\item The Henry-Draper Catalog (HD~209458)
\item The observatory or survey catalogs like the ones of CoRoT (e.g., CoRoT~9), Kepler (KIC~8462852) or TESS (TOI~986).
\end{enumerate}

In short, starting an exocomets' nomenclature designation with the name of its parent star indicates both its extrasolar nature and its observed location in the Universe. 

\subsubsection{Cometary nature}

Another critical aspect of a nomenclature scheme for exocomets is a designation of their cometary nature.  Here, we propose to add a single capital letter for this designation analogous to the nomenclature system applied to comets in the Solar System. 

The choice of the letter was discussed during the workshop on “Exocomets: Bridging our Understanding of Minor Bodies in Solar and Exoplanetary Systems”, held at the International Space Science Insitute (ISSI) in Bern in July 2024. There was a debate between the choice of the capital letter {\tt C} --- indicating the cometary nature of the designated object --- and the capital letter {\tt E} --- specifying the exocometary nature of the designated bodies. However, the use of the the letter {\tt E} for reference to the term ``evaporating" bodies as used in the acronym ``FEB" for Falling Evaporating Bodies is considered to be misleading. This is because sublimation and not evaporation is expected to be the main process at the origin of the out-gassing in exocomets. In addition, it has been considered that the letter {\tt E} for reference to the term ``exocomet" is not sufficently specific for these bodies. For example, it could also be used for other extrasolar objects like exo-asteroids or exo-moons. Therefore, the letter {\tt C} was decided to be the preferred choice. We adopt this choice in the final proposal for the nomenclature.

\subsubsection{Periodic nature and period,\\ and its exclusion from nomenclature}

Since some exocomets, like comets in the Solar System, can be periodic, and their periods may be determined in the near future \citep[as suggested for the transiting object discovered by][]{Kiefer_2017}, both their periodic nature and their measured period could be included among the parameters encoded in the nomenclature scheme. This would be analogous to the approach used for the periodic nature of Solar System comets, indicated by the capital letter {\tt P}, like in the comets 1P/Halley or 67P/Churyumov-Gerasimenko. 

However, the periodic nature of exocomets is not accessible to most observational facilities in most cases. Therefore, we propose to not include any indication of periodicity in the exocomets nomenclature.   

\subsubsection{Date of observation}

To individually characterize an observed exocomet, some specific information must be added to the nomenclature designation.  
The most straightforward approach to specify a given exocomet is to indicate the date of the discovery observation. This is typically performed for gamma ray bursts and gravitational waves events. \textit{Individual} exocomets discovered up until now have all been discovered via transit observations. Therefore for all these exocomets the date of observation is also the date of the transit characterizing the spatial position of the cometary body on its orbit.  

Most importantly, the indication of the date of observation also allows {\it a posteriori} the identification of multiple independent observations of the same object.

It is worth expanding upon the possible future circumstance where non-transiting exocomets are discovered. These hypothetical non-transiting exocomets may be discovered with, for example, direct imaging techniques. As we are dealing with individual bodies, the date of observation is also the date at which the position on the sky has been determined. As a result, for transiting and non-transiting exocomets, the date of the observation is always associated with some information on its position on its orbit. 

Note here that this presents similarities with the nomenclature of Solar System comets. Specifically, the time of the discovery is typically indicated in the designation by a combination of the year followed with a letter indicating the half-month and a number indicating the order of the discovery in the half-month (for example, C/2020 F3 (NEOWISE) was the third comet discovered the second half of the month of March 2020). However, such a temporal scheme is not applicable for exocomets. Specifically, in some exocometary systems like in $\beta$~Pictoris, many exocomets have been observed within the same half-month. Moreover, the order of the discovery does not automatically follow the chronological order of the observations. Therefore, we propose to not adopt the same scheme for exocomets. 
A more precise time indication than the one implemented for solar system comets is necessary for exocoments. Therefore, we advocate to include the date of the observation in the nomenclature in the form of year-month-date, in a similar way as done for gamma ray bursts.

The proposed format to indicate the date can be summarized by 
{\tt yyyymmdd}.

\subsubsection{Method of detection and physical properties,\\ and its exclusion from nomenclature}

The detection method (spectroscopy, photometry, and possibly in future imagery) could be included in the nomenclature scheme. 
This could be implemented via the addition of two letters like ``{\tt SP}" for spectroscopy, ``{\tt PH}" for photometry and ``{\tt IM}" for imagery. 

Moreover, the detection method may be conveyed indirectly with other information on the physical properties of the exocomets. 
For example, (i) transit depth, (ii) position angle, and (iii) radial velocity could be indicated for photometric, direct imaging and spectroscopic discovery respectively. Specifically for spectroscopy, the radial velocity information would enable identification of individual exocomets detected within the same spectrum. 
These further details may prove useful when the nomenclature constraints in the previous subsections are no longer sufficient to designate a single object. For instance, it is common to simultaneously detect several comets in the same spectrum of $\beta$~Pictoris \citep[][]{Kiefer_2014a}. In this case, the indication of the radial velocity at an accuracy of a few kilometers per second would be sufficient to distinguish several comets observed simultaneously.

We believe that the incorporation of too much information in the nomenclature scheme is not practical. 
Therefore, we propose to not add any information on the detection and/or the physical properties in the nomenclature scheme until it is required for individual exocomet identification. 

 \subsection{An additional optional letter for exocomets of the same day}

In some cases multiple exocomets are observed on the same day, which would be degenerate with the proposed nomenclature scheme so far. Therefore, we propose to differentiate these cases by simply adding a single lower case letter. This optional letter is to be added immediately following observation date in order to uniquely identify the designated exocomet.

This scheme is similar to the nomenclature scheme for multiplanetary extrasolar systems. In these cases the lower case letters ``b", ``c", ``d", etc. are used to indicate individual planets orbiting the same star, independently of their order in period or semi-major axis. The addition of an optional letter immediately following the date is also used in the nomenclature of gamma ray bursts to ensure that bursts occurring on the same day can be clearly distinguished from one another.

Here we propose to start with the letter ``a", and continue in the alphabetical order. As for exoplanets, we propose to let the discoverers select the order of individual comets discovered on the same date. 

Our proposed nomenclature could result in confusing designations in the unlikely event that multiple comets orbiting the same star are independently and contemporaneously discovered by different teams on the same day. This could be resolved by including a fractional day in the discovery date used for the designation if such a situation were ever to occur.
 
\section{Proposed naming convention}
\label{Proposed naming convention}

\subsection{The nomenclature scheme}

\subsubsection{{\tt Name-of-the-star}}
The first part of the nomenclature designation is based on the name of the star which the exocomet is associated with. Multiple stellar catalog names are possible (see Sect.~\ref{Name of the star}). The discoverer should choose one of these, as is done for exoplanets. We recommend selecting the name of the star following the priority order in the stellar catalogs as described in Sect.~\ref{Name of the star}. 

\subsubsection{Exocomet designation with {\tt C}}
In analogy with the convention used in Solar System comets nomenclature, it is proposed to use the capital letter ``{\tt C}" to specify the cometary nature of the designated bodies.

\begin{figure}
\centering
\includegraphics[width=0.9\columnwidth]{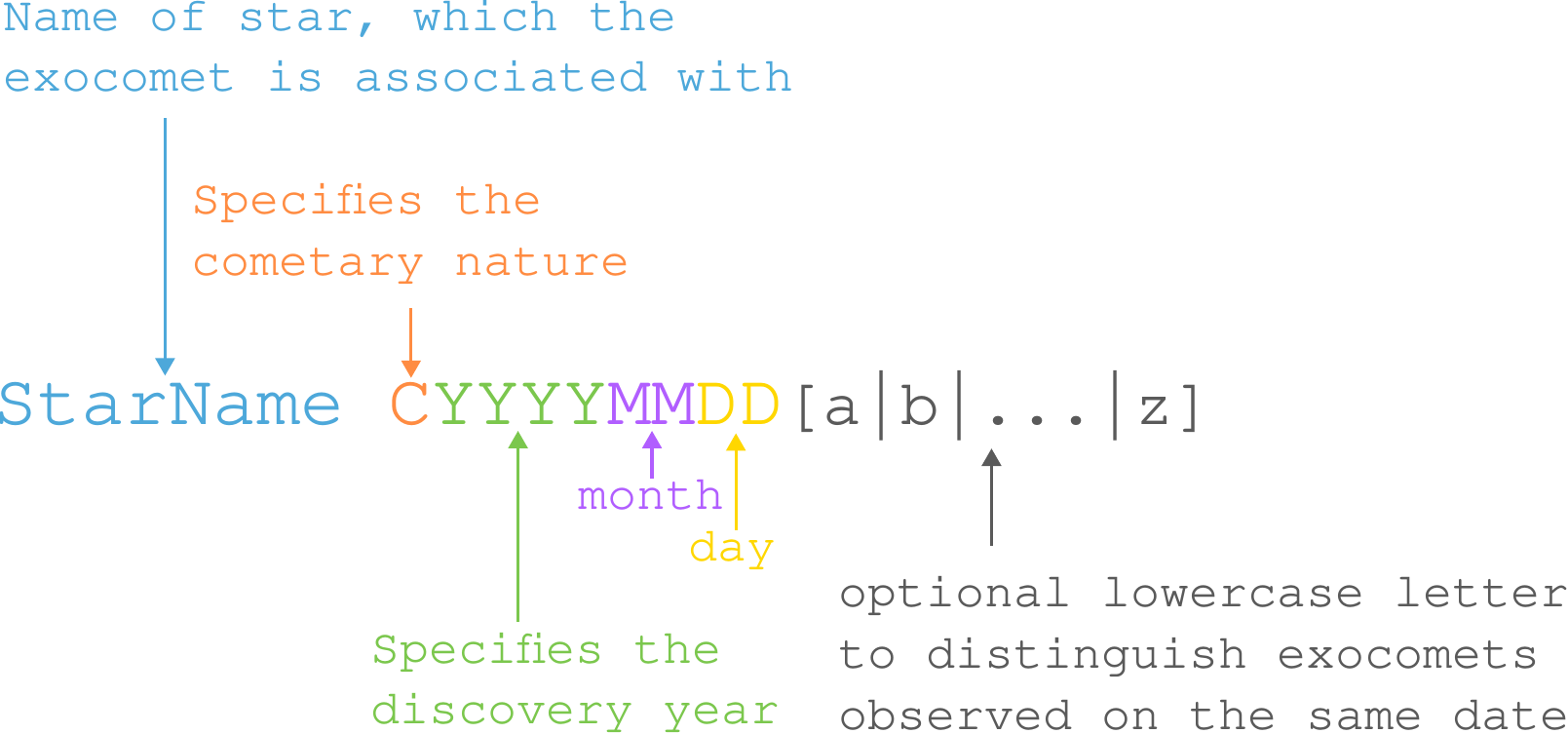}
 \caption{Structure of the nomenclature with each of the elements highlighted with a description.}
\label{fig:nomenclature}
\end{figure}

\subsubsection{Date of observation in {\tt yyyymmdd} format}

The date of the observation is then added in the {\tt yyyymmdd} format, 
without any space between the preceeding {\tt C} and the date.

\subsubsection{Optional differentiating lower case letter {\tt [a|b|\ldots|z]}}

If necessary to distinguish between several comets observed on the same date, an optional lower case letter can be added immediately following the date of observation without a space. This results in a designation consisting of eight numbers plus one optional letter for the information on the date of observation in the format {\tt yyyymmdd[a|b|\ldots|z]}.\\

A graphical summary of the proposed nomenclature naming structure is shown in Fig.~\ref{fig:nomenclature}.

\section{Examples}
\label{Examples}

Here we give some examples of use of the proposed nomenclature for a few exocomets. 
The proposed designations of the corresponding exocomets are summarized in Table~\ref{table:names}.
\subsection{$\beta$\,Pic\ C19841113}

The first exocomets identified as such are the ones described in the \cite{Ferlet_1987} paper. It seems that the first exocomet that can be clearly identified in the Figure~1 of this paper is the exocomet which we can designate by ``{\tt $\beta$\,Pic\ C19841113}".

\subsection{$\beta$\,Pic\ C19971206}

The exocomet that transited $\beta$~Pictoris in December 6, 1997 at a redshift of about 30\,km/s has been observed with the Hubble Space Telescope with the widest wavelength coverage ever obtained \citep[][]{Lagrange_1997}.
This comet has been analyzed in detail in 
\cite{Vrignaud_2024} and can now be designated by ``{\tt $\beta$\,Pic\ C19971206}".

\subsection{HD\ 172555\ C20110611}

In 2014, \cite{Kiefer_2014b} identified transits of four exocomets {\emph{in both lines}} of the Ca\,{\sc ii} doublet of the young star HD\,172555. 
One of those comets, seen with a small but significant blueshift of -2\,km/s can now be designated by 
``{\tt HD\ 172555\ C20110611}".

\subsection{$\beta$\,Pic C20091027a, C20091027b, C20091027c, C20091027d, C20091027e, and C20091027f}

\cite{Kiefer_2014a} made a statistical analysis of 1000~spectra of $\beta$\,Pic obtained between 2003 and 2011 with the HARPS spectrograph.
Figure~1 in that work presents a $\beta$\,Pic spectrum that contains not less than 6~transiting exocomets. These data were obtained on October 27, 2009, and show exocomets with radial velocities from -50\,km/s to +100\,km/s. 

These six exocomets can now be designated by 
"{\tt $\beta$\,Pic C20091027a}", "{\tt $\beta$\,Pic C20091027b}", "{\tt $\beta$\,Pic C20091027c}", "{\tt $\beta$\,Pic C20091027d}", "{\tt $\beta$\,Pic C20091027e}", and "{\tt $\beta$\,Pic C20091027f}".

\subsection{KIC\ 846285\ C20090805}

In the Kepler photometry data of the star KIC\ 846285 \citep[also nicknamed Tabby's star,][]{Boyajian_2016}, 
\cite{Kiefer_2017} discovered a repeated event with an absorption depth of 1000~ppm. These photometric variations can be interpreted as the first discovery of an exocomet photometric transit, or more probably a string of transiting exocomets. This exocometary string can now be designated by 
``{\tt KIC\ 846285\ C20090805}".

\subsection{KIC\ 3542116\ C20120522}
The first clear discoveries of single exocomet photometric transits are those made by \cite{Rappaport_2018} using visual inspection of the Kepler data. Six transits of KIC~3542116 and one transit in front of KIC~11084727 have been identified with light curves presenting the peculiar exocomet asymmetric shape as predicted by \cite{Lecavelier_1999}. The deepest transit occurred on May 22, 2012 with a transit depth of $1.5\times 10^{-3}$. 
The corresponding exocomet can now be designated by ``{\tt KIC\ 3542116\ C20120522}".

\subsection{$\beta$\,Pic\ C20181102}

Finally, the deepest and cleanest exocometary transit light curve has been discovered by \cite{Zieba_2019} in the TESS observations of $\beta$~Pic. The transit depth measured on the November 2, 2018 reached $2\times 10^{-3}$. 
The corresponding exocomet can now be designated by ``{\tt $\beta$\,Pic\ C20181102}".

\begin{table}
\caption{Proposed naming convention applied to the few examples given in Sect.~\ref{Examples}. 
}
\label{table:names}
\footnotesize
\begin{tabular}{ccc l@{}c@{}lc c c} 
Proposed Designation&Star&Reference \\ 
 {\tt $\beta$\,Pic\ C19841113} &$\beta$\,Pic&\cite{Ferlet_1987}\\ 
 {\tt $\beta$\,Pic\ C19971206}&$\beta$\,Pic&\cite{Lagrange_1997}\\
 {\tt HD\ 172555\ C20110611}&HD\ 172555&\citet{Kiefer_2014a}\\
  {\tt $\beta$\,Pic C20091027a}&$\beta$\,Pic&\cite{Kiefer_2014b}\\ 
 {\tt $\beta$\,Pic C20091027b}&$\beta$\,Pic&\cite{Kiefer_2014b}\\ 
  {\tt $\beta$\,Pic C20091027c}&$\beta$\,Pic&\cite{Kiefer_2014b}\\ 
    {\tt $\beta$\,Pic C20091027d}&$\beta$\,Pic&\cite{Kiefer_2014b}\\
{\tt $\beta$\,Pic C20091027e}&$\beta$\,Pic&\cite{Kiefer_2014b}\\ 
{\tt $\beta$\,Pic C20091027f}&$\beta$\,Pic&\cite{Kiefer_2014b}\\ 
{\tt KIC\ 846285\ C20090805}&KIC\ 846285&\cite{Kiefer_2017}\\ 
{\tt KIC\ 3542116\ C20120522}&KIC\ 3542116&\cite{Rappaport_2018}\\ 
{\tt $\beta$\,Pic\ C20181102}&$\beta$\,Pic&\cite{Zieba_2019}\\ 
\hline
\end{tabular}
\end{table}

\section{Conclusion}
\label{Conclusion}

The exocomet nomenclature described here is the result of discussions that occurred during the workshop on ``Exocomets: Bridging our Understanding of Minor Bodies in Solar and Exoplanetary Systems'' at ISSI, Bern, in July 2024. We encourage researchers to implement this proposed notation. It is useful to have a common designation of the same bodies that can be detected, analyzed and characterized in different studies.  
\\

This nomenclature is only a proposal that may be improved in near future. We hope that this nomenclature (or an improved version of it) will be adopted by the 
Working Group on Small Body Nomenclature (WGSBN) of the International Astronomical Union (IAU), and will be used worldwide in all studies mentioning individual exocomets.
Exocomets studies are now fully considered as part of the exoplanetology science. With new ground-based (e.g., ELT) and space facilities (e.g., PLATO) soon to come, this field will attract increasing interest. It is time to adopt a nomenclature to give the discovered 
exocomets the full existence as named astronomical bodies.

\backmatter

\bmhead{Acknowledgements}

We gratefully acknowledge support by the International Space Science Insitute, ISSI, Bern, for supporting and hosting the workshop on ``Exocomets: Bridging our Understanding of Minor Bodies in Solar and Exoplanetary Systems'', during which this work was initiated in July 2024.

AL thanks the support from the Centre National des Etudes Spatiale (CNES). 

D.Z.S. is supported by an NSF Astronomy and Astrophysics Postdoctoral Fellowship under award AST-2303553. This research award is partially funded by a generous gift of Charles Simonyi to the NSF Division of Astronomical Sciences. The award is made in recognition of significant contributions to Rubin Observatory’s Legacy Survey of Space and Time. 
\section*{Declarations}
The authors declare that they have no competing interests.

\bibliography{sn-bibliography}

\end{document}